# A Hydrodynamical Analysis of the Steady-State Shock Model

A.Preite Martinez*

INAF/IASF-Roma, Via del Fosso del Cavaliere 100, I-00133 Roma, Italy,  and
Observatoire de Strasbourg, 11 rue de l'Université, F-67000 Strasbourg, France

Abstract: In this article some of the hydrodynamical (HD) aspects of steady shocks as described by the steady-state shock model are reviewed and discussed. It is found that, at least in some of the contexts in which the steady-state model is used, the steady-state assumption cannot be satisfied. Moreover, the main result of the present work is that even if the assumptions on steadiness and on the geometry are fully satisfied, serious limitations in the application of the model are found: (i) in the absence of down-stream boundary conditions the model is not related to the physical process(es) that originate the shock, (ii) matter shocked during the presumed phase of steadiness of the shock is not hydrodynamically interacting with previously shocked matter, and (iii) the steady-state model assumes that the flow is stable against perturbations. Furthermore, even if boundary conditions were assumed, the link between the steady model and the astrophysical context would not be strictly speaking the correct HD link. Time-dependent HD computations in different astrophysical contexts (e.g. SNRs and molecular shocks) show that the steady-state approximation is inadequate to describe these post-shock structures. Based on the HD limitations of the steady-state model, it is advised that the model be used to describe post-shock structures only in those astrophysical contexts where full time-dependent HD models have already positively tested the steadiness of the flow. Alternatively, it is suggested to replace the steady-state model either with time-dependent HD models, or with less problematic approximations.

## 1. INTRODUCTION

In the sixties and the early seventies a simple model was developed to describe the effect of shocks in a variety of astrophysical contexts: from stellar atmospheres [1], to cloud collisions [2-4], supernova remnants (SNRs) [5-9], Herbig-Haro objects [8, 10] and AGNs [11, 12].
Work based on this model has been, and it is still, going on. A rather incomplete (and subjective) selection of more recent papers can include further works on AGNs [13-17], SNRs [18-22], proto-PN and PN [23, 24], proto-galactic matter [25, 26], and pulsating stellar atmospheres [27, 28].
The model took the name of "steady-state shock", from the assumption that the flow is steady, or "steady-state radiative shock", because the flow is cooling and radiating away part of its energy, or "plane-parallel steady shock", from the assumed geometry.
The steady-state structure is used in connection with a photo-ionization code, in order to generate emission in all sort of bands and lines, and try to fit the emissions observed in astrophysical shocks. In this article only the hydrodynamical aspects will be discussed.

In the following section some of the HD aspects of the steady-state model, and the assumptions used, are analyzed in detail. Analytical expressions are derived to express the cooling time-scale and the HD time-scale as a function of the shock velocity (one of the input parameters of the model). These expressions are used to test the steady-state assumption in particular cases.
To have a better insight of the steady-state model and to understand under which circumstances this model can be used as an approximate description of astrophysical post-shock structures, steady-state computations have been compared with time-dependent HD models, in particular in the context of SNRs [29, 30] and, more recently, in the context of molecular shocks [31-33].
Another important aspect is the stability of steady shocks against small perturbations. In the case of colliding supersonic flows, Walder & Folini (1998) [34] and Folini & Walder (2006) [35] use 2-D high-resolution numerical simulations to describe the collision of flows up to Mach-number 90. They show that strong turbulent zones form behind the shocks. In the case of SNRs, "it seems clear that steady radiative shock models are of no relevance for shock velocities" above 150 km s$^{-1}$ [36] or above 90 km s$^{-1}$ [37]. Lesaffre *et al* (2005) [31] find a molecular instability for a short range in shock velocity (20-25 km s$^{-1}$) in molecular shocks. Smith & Rosen (2003) [38] also show that moderately fast shocks in dense molecular clouds with low transverse magnetic fields are likely to be unstable up to shock velocities of 70 (or possibly 80) km s$^{-1}$. On the other hand, Innes (1992) [39] showed that sufficiently strong magnetic fields (e.g.: 9 µG) could damp the formation of instabilities in the flow of shocks up to 175 km s$^{-1}$.

A discussion of some of the above comparisons is given in the following sections. In the conclusions, the limitations found in the HD analysis of the steady-state shock model are summarized, and other solutions are suggested.

## 2. THE HYDRODYNAMICAL PROBLEM

In order to derive the HD structure of a generic flow, one needs to integrate a set of partial differential equations (PDE; e.g. the Euler's equations) with the appropriate up-stream and down-stream boundary conditions and initial conditions (i.e.: the HD structure of the flow at time $t = 0$). In the presence of a shock, the up-stream boundary conditions are given by the values of the physical variables of the pre-shocked medium. At the shock front the flow variables are subject to abrupt changes, but there are simple exact relations relating the flow values ahead of the shock with those immediately following the shock discontinuity.

These relations are called the Rankine-Hugoniot (RH) conditions (see for instance the formulation given in Landau & Lifshitz,1971) [40]. They are derived from the standard system of hydrodynamical PDEs, assuming that the crossing of the discontinuity is instantaneous. Then all the time derivatives vanish in the rest frame of the shock front, as they would do if the whole physical system were independent of time: but it is worth stressing that if these two cases are mathematically equivalent, they have a completely different physical meaning.

The system of PDEs becomes a system of algebraic eq.s that impose the conservation across the discontinuity of the fluxes of matter, impulsion, and energy. The eq.s are written in a reference frame moving with the shock. The RH relations are exact, and they assure the correct increase of the entropy for matter crossing the shock.

The down-stream boundary conditions depend on the context (piston, wall, adiabatic expanding sphere, energy input, etc.). The initial ($t = 0$) structure of the flow should also be given, and it could also depend on the context under study. The initial structure is of fundamental importance also because it sets the total energy of the system.

The steady-state model is an attempt to simplify the HD problem. It relies on the following basic assumptions:

(i) the flow is plane-parallel, "because the cooling region is so thin compared to the radius of curvature of the shell" [5], i.e. the radius of curvature of the shock is much larger than the typical linear scale of the post-shock region to be described. In order to apply the model to shocks generated by jets or outflows, more recently a new generation of models has been presented [13,16], where the shock surface is assumed to have the shape of a bow shock.

(ii) the flow is steady. There are different formulations in the papers cited above to say that the flow into the shock does not vary in the time required for the development of cooling in the post-shock region. Probably the best one is: the flow parameters "remain constant during the time required for gas to cool from the post-shock temperature to a temperature low enough that optical emission ceases" [8]. The flow parameters are the shock velocity $V_s$ and the physical and chemical properties of the ambient pre-shocked medium, as the number density $n_0$. Or: "The restriction to steady-state shocks is physically justified as long as the cooling time is short compared to the times for the shock velocity and the pre-shock conditions of the gas to change" [25].

With these assumptions, the computation of a steady-state structure is much simpler than a time-dependent HD computation.

The model makes use of the Rankine-Hugoniot eq.s to define the up-stream boundary conditions. Then, a critical point: "since the jump conditions relate quantities at an arbitrary point behind the shock front to those ahead of the shock front" [4] these conditions are used – together with a time-dependent energy equation accounting for cooling – to compute not only the effect of the instantaneous crossing of the shock discontinuity, but also to describe the entire down-stream flow.

The system of steady-state eq.s is completed with the equations describing the ionization structure of the flow. So eventually the original system of PDEs is transformed into a system of ordinary differential eq.s (ODEs). These eq.s state that some properties of the gas (as the flux of specific mass, momentum and energy, including radiative) are constant along the flow. The eq.s used to compute steady-state shock structures are described in most of the papers cited above. The model doesn't need an initial structure of the flow. The derived structure can be considered – a posteriori – an initial condition, if the flow is indeed steady.

With the exception of computations of steady-state shocks driven into the atmosphere of pulsating stars by the pulsating interior of the star, acting as a piston [1], the structure of the flow is always computed without using down-stream boundary conditions.

Indeed, the structure of a steady-state flow is computed starting from the up-stream boundary conditions, following a parcel of gas as it goes through a stationary shock, flows beyond the shock, and cools. It is worth noting that if the flow is steady as assumed, the spatial structure of the flow at a given time corresponds to the temporal evolution of a single test parcel of gas during the integration time $t_{end}$ of the model.

One could argue that the system of these so-called steady-state eq.s is no longer composed of PDEs, so the initial structure and the down-stream conditions are not formally necessary. But it will be shown in the discussion below that these conditions (in particular the down-stream conditions) are necessary to relate the flow to the physical context the model is applied to.

Furthermore, it will be shown in the following that, even when the two main assumptions of the model (on steadiness and on geometry) are satisfied, the steady-state model shows serious hydrodynamical limitations.

## 2.1. The steady-state assumption

A flow is called steady when all the time derivatives of the flow field vanish. For instance, one should find that the velocity of the fluid at a given point (the Eulerian velocity) remains constant with respect to time. It is not easy to find in astronomy post-shock flows that globally satisfy, even approximately, this condition. One can imagine a time-dependent HD set-up that, described by the usual set of PDEs and with a careful choice of boundary and initial conditions, can evolve into a stationary flow. In particular cases a change of the reference frame could transform a flow from unsteady to steady or, at least, locally steady.

Lesaffre *et al.* (2004) [32] find in their time-dependent HD computations that there can be limited regions of a flow where, by adopting a particular frame of reference, "none of the variables is changing in time". These regions are then called by the authors in a quasi-steady state. Note though that these (quasi) steady flows are obtained by integrating a system of PDEs, not by the steady-state eq.s, so they won't suffer of the limitations discussed in the following sections.

The steady-state model assumes that the cooling time $\tau_c$ of shocked matter is much shorter than the time scale $\tau_{HD}$ of significant variations of HD quantities affecting cooling (e.g. $V_s$ or the post-shock temperature).

Only an analysis of the astrophysical context can tell if the assumption that $\tau_c \ll \tau_{HD}$ is physically reasonable or not. Such an analysis is necessary because, by construction, a steady-state calculation will always produce a cold dense region, independently of the flow parameters (e.g.: the shock velocity) and of the astrophysical context.

There are cases in which the assumption that the cooling time is much shorter than the HD time can be easily verified. The cooling time $\tau_c$ can be defined as the e-folding time $\tau_c = \varepsilon/(d\varepsilon/dt)$, where $\varepsilon$ is the internal energy, $d\varepsilon/dt = n_e \Lambda$, and $\Lambda$ is the cooling rate of the plasma. Expressing $\Lambda$ in units of $10^{-22}$ erg s$^{-1}$ cm$^{-3}$, $T$ in Kelvin, $\tau_c$ in years and $\gamma = 5/3$, one gets

$\tau_c \sim 1.37 \times 10^{-2} \, T / (n_0 \, \Lambda)$.

The cooling rate $\Lambda$ is of course a function of the temperature, of the ionization structure, and of the composition of the cooling plasma. If a solar composition and a steady-state ionization structure are assumed, $\Lambda$ can be expressed as a function of $T$ using the power law fits of Taylor & Dyson (1992) [13]. Then, using the Rankine-Hugoniot conditions and assuming a strong shock, the post-shock temperature $T$ can be expressed in terms of the more handy input parameter $V_s$. Eventually one gets

$$\tau_c \sim 390 \, x^\alpha / n_0 \quad \text{yr} \qquad (1)$$

where $x = V_s / (103.4$ km s$^{-1})$, and $\alpha = 3.52$ or $\alpha = -0.7$ for $x>1$ or $x \leq 1$, respectively. Note that if the integration time $t_{end}$ of the steady-state model is different from $\tau_c$ the steady-state condition should be tested for $t_{end}$, i.e.: $t_{end} \ll \tau_{HD}$.

The HD time scale can be defined as the time required for a variation of a factor $k$ in the shock velocity $V_s$, which would correspond to a variation of a factor $k^2$ in the temperature just behind the shock. So

$$\tau_{HD} = \ln(k) \, | \, dt / d \ln(V_s(t)) \, | \qquad (2)$$

There is at least one physical context in which $\tau_{HD}$ can be expressed in analytical form: supernova remnants in the adiabatic phase. In this case $V_s \sim t^{-3/5}$, so that from eq.(2) we have $\tau_{HD} = \ln(k) \, t / 0.6$, or $\tau_{HD} \sim 0.16 \, t$ for $k = 1.1$ (i.e. for a 10% variation of $V_s$. Note that the choice of $k$ is related to the astrophysical problem under study. While $k=1.1$ can be a reasonable value for SNRs, in other contexts where the post-shock structure strongly depends on $V_s$ [33] the value of $k$ can be closer to 1.0). In the radiative phase $V_s$ can become a much steeper function of time and $\tau_{HD} \sim 0.02 \, t$ or even smaller.

The steady-state condition $\tau_c \ll \tau_{HD}$ can be transformed in a condition on $V_s$. To do so, also $\tau_{HD}$ has to be expressed in terms of the shock velocity. Then, inverting the well known relation $V_s \sim t^{-3/5}$ for adiabatic remnants, eq.(2) becomes

$$\tau_{HD} \sim 5.05 \times 10^7 \, (E_{51}/n_0)^{1/3} \, V_s^{-5/3} \quad \text{yr} \qquad (3)$$

where $E_{51}$ is the total energy of the remnant in foe. Using $\tau_c$ as expressed in eq.(1) and $\tau_{HD}$ from eq.(3), the steady-state condition $\tau_c / \tau_{HD} \ll 1$ can be rewritten as a condition on $V_s$:

$$V_s \ll 226 \, E_{51}^{0.064} \, n_0^{0.1285} \quad \text{km s}^{-1} \qquad (4)$$

Because of the very low dependence on total energy and ambient density, it can be claimed that shocks in SNRs with $V_s > \sim 100\text{-}150$ km s$^{-1}$ cannot be steady. If we set both $E_{51}$ and $n_0$ equal to 1, this finding can be cross-checked. The age corresponding to $V_s = 226$ km s$^{-1}$ is 37,600 yr. But from [41] (see their eq.7) a remnant with such a shock velocity is already well into the radiative phase. Different values of $E_{51}$ or $n_0$ will not change the result: the steady-state assumption cannot be satisfied in remnants in the adiabatic phase.

In SNRs in the radiative phase the situation can be different: $\tau_{HD}$ increases because of the factor $t$ (the age of the remnant) even if the numerical factor decreases as discussed previously. The cooling time can be reduced by increasing the ambient density $n_0$. Thus there can be radiative SNRs with high $n_0$ (e.g. $n_0 \sim 6$ cm$^{-3}$ as estimated by Cox (1972) [5] in the Cygnus loop) in which $\tau_c < \tau_{HD}$. A cautious statement is then that in the presence of a high $n_0$ the steady-state assumption can be satisfied in SNRs in the radiative phase. But, as it will be shown in the following, the limitations of the model are elsewhere. Let's now consider truncated steady shocks with a shock velocity not satisfying condition (4). For example, let's take a $V_s = 500$ km s$^{-1}$ steady-state shock model from Dopita & Sutherland (1996) [15] and apply it to a SNR expanding into an ISM with number density $n_0 = 1$. We know from time-dependent HD calculations that a remnant with such a shock velocity is not yet in the radiative phase. Using the analytical relations valid for the adiabatic phase, we can evaluate its age $t \sim 10,000$ yr and the post-shock temperature $T_s \sim 3.5 \times 10^6$ K. In this case $\tau_{HD} \sim 1,600$ yr. In the steady-state model, a cold shell should form after $\tau_c \sim 100,000$ yr and the integration time is of the order of 300,000 yr [15]. Both the cooling time and the integration time $t_{end}$ of the model are not only in contrast with the steady-state assumption, but also much longer than the age of the remnant.

One could then truncate the integration of the model at $t_{end} < \tau_{HD}$, presuming that the truncated steady-state flow structure might represent a reasonable approximation of a time-dependent HD structure. Even so, the steady-state eq.s and the time-dependent eq.s produce different behaviours for the flow variables: constant values (those just behind the shock discontinuity) and the self-similar solution (e.g. the one referring to spherical geometry), respectively.

## 2.2. The geometry

The other basic assumption of steady-state shock models is that the flow is plane-parallel. This assumption is reasonable in most cases (e.g.: old SNRs, cloud collisions), but it fails when the geometry of the shock surface is far from being plane, as in young SNRs and in bow shocks generated by jets (e.g.: AGNs, H-H objects, proto-PNe). To overcome this problem, Taylor *et al* (1992) [13] and Ferruit *et al* (1997) [16] developed a model of bow-shock for AGNs in which the shock surface is a paraboloid of the 4th order. Each parcel of ambient gas crossing the bow surface is then followed as a separate steady-state flow. Recent applications of this model concerns not only AGNs but also bow shocks around condensation knots in PN [23] and proto-PN [24].

In these models the description of the geometry seems to be rather accurate. Unfortunately, the increased accuracy of the geometry of the problem translates into a serious drawback in the HD treatment of the flow. The serious limitation resides in the different steady-state flows forced to share the same volume (actually, the bow surface) without being allowed to hydrodynamically interact.

The condition on the steadiness of the flow has an influence on its spatial extent. For a flow to be described by the steady-state eq.s, no perturbation should reach the flow while the parcel is travelling down-stream and is cooling. In this context, a perturbation is any interaction of the flow with the hydrodynamical information carried by a different flow (steady or not steady) in different physical conditions. This condition defines a surface surrounding the flow in all its spatial extent that must be at least – to be on the safe side – at a distance $c_2\, t_{end}$ from the flow, where $c_2$ is the (highest) sound speed in the flow: $c_2^2 = \gamma\, p_2/\rho_2 \approx V_s^2$. This means that the steady-state assumption must be satisfied in a cylindrical region of height $V_s\, t_{end}$ and area of its base $(V_s\, t_{end})^2$. Moreover, the input parameters of the different flows crossing the base surface need to be uniform across the entire base surface. Otherwise different parameters will yield different, and then interacting, steady flows. This means that the above geometrical considerations add a *homogeneity condition* to the steadiness assumption.

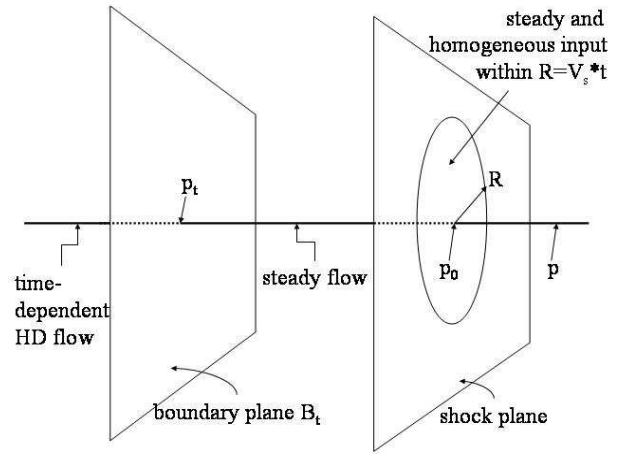

Fig.1. In the figure we schematically show the stream line defined by the test particle $p$, as it crosses the up-stream boundary plane (the shock plane) at $t=0$ and the down-stream boundary plane B at time $t$. The homogeneity condition must apply in the region of radius $R=V_s\, t$ surrounding the stream line $p$.

## 2.3. The boundary and initial conditions

Let's consider now how the absence of one or more among the two boundary conditions and the initial structure of the flow influence the link between the astrophysical context and a structure computed with the steady-state model and assumptions.

Let's assume that the two basic assumptions of steady-state models (on steadiness and geometry) are satisfied in a particular case. It will be also assumed (as it is the case for practically all models computed since 1972 [5]) that there is a preferential direction for the integration of the system of steady-state ODEs, namely starting from the up-stream boundary conditions.

This choice is of course a very practical one, because it allows to parameterize the flows with input values that are often observationally constrained, and whose meaning is subjectively easy to understand (as the shock velocity and the density of the pre-shocked medium).

On the other hand, it is in principle possible to build the steady-state flow starting from a set of down-stream conditions, at the point where the time-dependent flow is supposed to become steady. This is possible but certainly not very practical in most cases, because it requires the knowledge of the time-dependent HD conditions further down-stream, and the parameterization would be done in terms of values of the HD variables (pressure, density, velocity at this down-stream boundary called $B_t$ in Fig.1) not easy to locate and observationally constrain. The advantage would be a clear link between the astrophysical context and the shock, through the conditions at the down-stream boundary and the propagation of the steady-state fluxes of specific energy, momentum and mass up to the shock.

Keeping this in mind, let's compare the evolution of a test parcel of gas, as it goes through the shock and it traces the HD structure of the post-shock region, as described by a time-dependent approach and by the steady-state approach.

Let's first consider the time-dependent approach. We start with the test parcel of matter $p$ rushing (in the reference frame of the shock) at velocity $V_s$ towards a shock. The parcel is not aware of the approaching shock, because the sound speed is lower than $V_s$. Nor *a fortiori* of the astrophysical nature of the shock. Then the parcel hits the shock, and its status is suddenly changed according to the RH conditions. Still the parcel doesn't know which is the context that has generated the shock, i.e. it doesn't know if it is entering a SNR, a high-velocity cloud around an AGN, or else. This is because the RH conditions apply instantaneously, and to communicate the nature of the HD flow generating the shock to the parcel during this (formally instantaneous) phase, waves propagating at infinite speed would be necessary. As soon as the formalities of the passage of the shock accomplished, the parcel enters the post-shock region. There it realizes which is the astrophysical context that has generated the shock, because it is entering a "sonic" region, i.e. the HD information is carried around by waves propagating at the speed of sound. So for the same given values of the input parameters, the corresponding HD structures of the post-shock flow traced by our test parcel will in principle differ from context to context. Because, once emerged from the shock, the test parcel will interact with previously shocked matter.

Let's now compute the fate of the same test parcel of matter $p$ hitting the shock at time $t=0$ with velocity $V_s$, as described by the steady-state eq.s. The test parcel will see its parameters being modified according to the RH conditions, as in the time-dependent case. But as it starts receding from the shock, it is still un-aware of the nature of the shock. Moreover, it will trace a structure that will only be a function of the input pre-shock, up-stream conditions. So at given values of the input parameters will always correspond the same post-shock structure. This is because the steady-state eq.s are built from the RH conditions. In other words, the test parcel will never know if it is entering the HD structure of a SNR or of a high-velocity cloud or else, because formally it will never emerge from the shock into the "sonic" region! And it will never interact exchanging sonic waves with previously shocked matter.

Formally, as already pointed out, the only form of interaction that the system of steady-state ODEs can assure is that the flux of some specific quantities is constant from the start of the integration at the shock plane, to the down-stream plane $B_t$ at the end of the integration.

The post-shock region down-stream w.r.t. the steady structure traced by our test parcel (i.e. down-stream w.r.t. $B_t$) is occupied by matter shocked under conditions that were different from those assumed in the steady-state model, e.g. implying different fluxes of energy, momentum and mass. The shocking conditions must have been different, otherwise the region would be part of the steady-state structure, and one could repeat this reasoning further down-stream.

In summary, the assumptions that are made to integrate the steady-state eq.s imply that *matter shocked during the presumed phase of steadiness of the shock is not hydrodynamically interacting with matter shocked in previous phases* (non-steady or steady with different input parameters). This means isolating the steady structure from the physical context that generated the shock.

This is true for two basic formal reasons: because (i) waves are not a solution of the system of steady-state eq.s, and (ii) for the model the only form of interaction is the propagation within the flow of the constant fluxes of the specific mass, momentum and total energy, and this information is not enough to fully link

(at the down-stream boundary) the steady flow to the time-dependent flow of previously shocked matter.

Note that the above implicit assumption is also valid for truncated steady structures: in these cases the problem is just displaced further down-stream w.r.t the truncated end of the flow.

Let's now consider the effect of a missing initial status of the structure. It can be noted that the model assumes that, because the flow is in steady-state, the calculated structure of the flow is – a posteriori – the initial flow status. So a particle crossing the shock at any time $t$ will always find a pre-existing steady structure. Even at time $t=0$. But presuming this, means to arbitrarily extend the interval of validity of the assumption that the flow is steady.

As said in Sect.2, the initial structure also sets the total energy of the system. Energy is a delicate problem for steady-state shock models. The total energy of the system is unknown because the initial structure of the flow is unknown (not given), and because the spatial extent in the plane of the shock is un-determined.

On the other hand, the specific (per unit surface of the shock plane) total energy can be computed. It is the result of the conversion of the kinetic energy of the un-shocked medium at the shock discontinuity. In steady-state models the whole flow structure – not only the effect of the shock discontinuity – is computed in the shock co-moving reference frame. As a consequence, the ambient medium is assumed to be an un-limited energy reservoir, and the (specific) total energy of the shocked flow is a linear function of the integration time of the model.

In time-dependent HD computations the shock works at the expenses of the already shocked gas, i.e. the conditions of the down-stream flow (for a given ambient medium) set the strength of the shock and then the amount of energy that is transferred to the gas just shocked. In the steady-state model, with the ODEs integrated starting from up-stream, energy is drawn from the un-shocked medium up-stream and it is flowing with matter down-stream.

It could be argued that discussing total energy for a steady-state flow is meaningless, not only because of the geometry assumed, but because the flow describes only a part (the radiative one) of the astrophysical system under study. This argument would be true if the flow were described as an open thermodynamical system, able to exchange energy (heat and work) and matter with its environment. On the contrary, the steady flow is described by eq.s that only imply the propagation within the flow of the constant fluxes of the specific mass, momentum and total energy.

Even if the assumptions on steadiness and on the geometry are satisfied, the use of the so-called steady-state eq.s and the absence of down-stream boundary conditions are a serious limitation of the model. All the HD information (e.g.: momentum and energy) necessary to keep the shock running at constant speed while compressing and heating the ambient material, is coming from down-stream, and it is set by the down-stream boundary conditions and the past HD history of the flow.

## 3. COMPARISON WITH TIME-DEPENDENT HD MODELS

Soon after the first works based on the use of steady-state shocks a number of time-dependent hydrodynamical models were published. Among these, only two [29, 30] explicitly compared the result of the time-dependent HD computations with the assumptions of the steady-state model. The first comparison was made by Chevalier (1974) [29]. In this paper on spherically symmetric models of supernova remnants the author, commenting the results of his computations, explicitly confirmed the steady-state assumption with the following paragraph: "The time for gas to move from the shock to the recombination region was small compared with the time for the shock velocity to change significantly. Thus," as in Cox (1972) [5], the quantity $P = B_t^2/8\pi + \rho v^2 + nkT$ "is expected to be constant through this region. This was found to be approximately true in the integration".

In a subsequent paper explicitly devoted to the formation of a dense shell in SNRs in the radiative phase, Preite-Martinez (1981) [30] found that "the time for gas to move from the shock to the recombination region is not small compared to the time for the shock velocity to change significantly". Actually, the author found that during this time the shock velocity changed by about 40%, leading to a change in post-shock temperature of a factor of about 2. Moreover, the quantity $P$ defined above "that should be constant behind the shock, as a direct consequence of the assumptions of the steady-flow model" changed by more than one order of magnitude in the same region.

A technical explication of the discrepancy with Chevalier (1972) [29] can be found in Preite-Martinez (1981) [30].

There is a strong argument *against* steady-state flow computations in Chevalier's (1974) paper [29]. This can be found in point 7 of his conclusions: "Material accretes onto both sides of the dense shell", but the statement doesn't make reference to the much different behaviour of steady-state flows. Chevalier's conclusion is confirmed by Preite-Martinez (1981) [30] : "the cold dense region behind the shock is accreting matter also through its trailing edge, hence real HD computations find the thickness of the cooling region more than one order of magnitude wider" than steady-state computations.

Another test can be performed concerning the main assumption of steady-state computations: i.e. that the evolution with time of the HD and ionization conditions of matter flowing along the stream line hit by the shock at the given velocity $V_s$, is assumed to coincide with the spatial structure of the flow behind a shock whose velocity is $V_s$ [5, 8, 19]. According to the models of Preite-Martinez (1981) [30] the test is negative: the evolution with time of the flow parameters (density, velocity, pressure, etc.) of matter flowing along a stream line that crossed the shock at a given velocity $V_s$ is not even qualitatively similar to the spatial structure of the flow when the shock velocity is $V_s$. This is because the spatial structure of the flow is made up of all the stream lines that crossed the shock at previous times, and they are interacting because the flow is sonic.

The reason why both authors are not confirming the results of steady-state computations is better understandable in the light of what has been derived in Sect. 2.1. Both authors are modelling SNRs evolving in rather low ambient densities ($n_0 = 1$ and 0.1), where the steady-state assumption is not satisfied. Nonetheless, the comparison keeps its validity. The combined accretion of matter (from both the leading and the trailing edges) into the cooling region is a phenomenon that is related to the interaction of matter shocked at different times. The presence of this interaction is not related to the density of the ambient medium. Thus, the physical, dynamical and geometrical properties of the cooling region in time-dependent HD calculations of SNRs are always drastically different from those computed with the steady-state approximation.

More recently there has been some work [31, 32] on shocks in molecular clouds, were the authors test the local validity of the steady-state equations against the time-dependent evolution of molecular shocks. They find expressions to characterize the steadiness of the flow at any point of the flow. The authors find that there can be local regions in the flow that are "quasi-steady" and can be modelled by a truncated steady-state model. They derive methods for the reconstruction of time-dependent shocks using truncated steady-models. Those constructions will be meaningful only for shocks in which the quasi-steady state has been validated at all times. They warn though that "there is still no other way to assess the validity of the steady-state assumption than computing the time-dependent evolution with a fully HD code" [32]. Even though "quasi-steady shocks are not strictly speaking time-dependent shocks" [33], they are a good approximation to them. More simpler approximations as the "steady-state approximation badly fail to reproduce the behaviour of the shock" over long time periods because, as explicitly recognized long ago [30], the past history of the shock is essential in determining its present behaviour.

As summarized in the Introduction, many authors have built 1-D or 2-D time-dependent HD models and used them also to test the stability of post-shock structures in different astrophysical contexts. They all agree on the conclusion that fast shocks are unstable, then they cannot be steady. The notion of "fast" is of course context-dependent, but a general lower limit can be set in the range 100-150 km s$^{-1}$.

Nonetheless, models of steady-state shocks were, and still are, computed for shocks of much higher velocities: up to 200 km s$^{-1}$ [8, 42, 43], 300 km s$^{-1}$ [10, 24, 44], up to 500 km s$^{-1}$ [14], 700 km s$^{-1}$ [16], 790 km s$^{-1}$ [22], up to, and above, 1000 km s$^{-1}$ [12, 13, 45].

If we consider only the field of SNRs, it is worth noting that time-dependent HD simulations have been used since the appearance of the first steady-state model of a SNR. See for instance Cox (1972) [5] and the companion paper Cox (1972) [46]. More recently we can quote the works of Tenorio-Tagle *et al* (1991) [47], Cui and Cox (1992) [48], Blondin and Lundqvist (1993) [49], Chevalier and Dwarkadas (1995) [50], Borkowski *et al* (1996) [51], Truelove and McKee (1999) [52], and Dwarkadas (2005, 2007) [53, 54]. A HD comparison with recent works using steady-state models is practically impossible, because while papers based on time-dependent models mostly deal with the global properties and evolution of the remnant (even in environments where the input parameters, as the density, are far from constant), steady-state models are basically used as a tool to interpret spectroscopic data.

## CONCLUSIONS

Some of the HD aspects of steady shocks as described by the steady-state shock model have been reviewed and discussed in detail. First, the assumptions used in the model have been analysed. Analytical expressions have been derived in order to express the cooling and the HD time-scales as a function of the shock velocity. With the help of these expressions, the steady-state assumption can be tested in particular astrophysical contexts. It was found that the steady-state assumption is never verified in SNRs in the adiabatic phase, so that the steady-state model cannot be used to describe the HD structure of adiabatic remnants. In SNRs in the radiative phase, the steady-state assumption can be verified if the ambient density is high.

But even supposing that the fundamental assumptions on steadiness and geometry are satisfied, due to the choice of the equations that describe it and to the choice of a preferential direction of integrations, there are serious limitations in the application of the steady-state model to astrophysical shocks. Among other problems, it is found that (i) the steady-state model is not related to the physical process(es) that originate the shock, (ii) that matter shocked during the steady-state phase of the shock cannot interact with previously shocked matter as it would do in a time-dependent HD model, and (iii) the steady-state model assumes it remains stable against perturbations.

The time-dependent HD computations of SNRs [29, 30] and the recent work [33] in the field of molecular shocks, indicate that the steady-state approximation appears to be inadequate to describe post-shock structures in the tested astrophysical contexts. Although the stability analysis in different astrophysical contexts indicates that fast shocks are unstable (and then unsteady), models of steady-state shocks have been, and still are, computed for shocks of much higher velocities.

Based on the present analysis of the model, it is advised that the steady-state model be used to describe post-shock structures only in those astrophysical contexts where full time-dependent HD models have already positively tested the steadiness of the flow.

In view of the scientific work built around and above the steady-state model, concerning photo-ionization/dissociation and emission mechanisms, it is important to explore the conditions in which the steady-state model can be considered a reasonable approximation to time-dependent models.

Can the steady-state model be used as an approximation of time-dependent HD models, when the steady-state assumption is satisfied? In principle, yes. But the assumption has to be tested, and the degree of approximation has to be evaluated. One of the results of the present work has been to find an unambiguous way to test, at least in one astrophysical context, the validity of the steady-state assumption. Lesaffre et al (2004) [32] did it in the context of molecular shocks. It is crucial to verify the assumption in other contexts. A comparison with time-dependent models is also necessary to evaluate the degree of approximation on a case-by-case basis.

Moreover, it should be noted that in steady-state models the ionization structure of the shocked gas is correctly assumed as depending of time, and treated consequently. The experience with 1-D models shows that computing the time-dependent ionization structure for the most abundant elements is much more demanding in terms of computing power than the computation of the time-dependent HD structure. So, a reasonable suggestion can be to use (at least 1-D) time-dependent HD computations as input for the photo-ionization codes. With the advantage of linking the model(s) to the astrophysical context(s) and gaining a better understanding of the problem(s) under study.

*Acknowledgements.* I thank Prof. Jean Heyvaerts, Observatoire de Strasbourg, Dr. Françoise Genova, C.D.S. Strasbourg, and Dr. Roland Gredel, Max Planck Institute for Astronomy, Heidelberg, for a critical reading of the manuscript. I also thanks the referees for their useful comments and in particular Dr. P.Lesaffre whose help was very important in clarifying the text.

# REFERENCES

[1] Whitney CA, Skalafuris AJ. The structure of a shock front in atomic hydrogen. I. The effects of precursor radiation in the Lyman continuum. ApJ 1963; 138: 200.
[2] Field GB, Rather JDG, Aannestad PA, Orszag SA. Hydromagnetic shock waves and their infrared emission in HI regions. ApJ 1968; 151: 953.
[3] Shull JM, Hollenbach DJ. H2 cooling, dissociation, and infrared emission in shocked molecular clouds. ApJ 1978; 220: 525.
[4] Hollenbach D, McKee CF. Molecule formation and infrared emission in fast interstellar shocks. I. Physical processes. ApJS 1979; 41: 555.
[5] Cox DP. Theoretical structure and spectrum of a shock wave in the interstellar medium: the Cygnus loop. ApJ 1972; 178: 143.
[6] Dopita MA. Optical emission from shock waves. I. Abundances in N49. ApJ 1976; 209: 395.
[7] Dopita MA. Optical emission from shock waves. II. Diagnostic diagrams. ApJS 1977; 33: 437.
[8] Raymond JC. Shock waves in the interstellar medium. ApJS 1979; 39: 1.
[9] Shull JM, McKee CF. Theoretical models of interstellar shocks. I. Radiative transfer and UV precursors. ApJ 1979; 227: 131.
[10] Dopita MA, Binette L, Schwartz RD. The two-photon continuum in Herbig-Haro objects. ApJ 1982; 261: 183.
[11] Daltabuit E, Cox D. A plausible energy source and structure for quasi-stellar objects. ApJ 1972; 173: L13.
[12] Daltabuit E, MacAlpine GM, Cox DP. The structure and spectrum of a colliding-cloud system and its possible relationship to QSOs. ApJ 1978; 219: 372.
[13] Taylor D, Dyson JE, Axon DJ. Bowshocks and the formation of the narrow-line region of Seyfert galaxies. MNRAS 1992; 255: 351.
[14] Dopita MA, Sutherland RS. Spectral signature of fast shocks. II. Optical diagnostic diagrams. ApJ 1995; 455: 468.
[15] Dopita MA, Sutherland RS. Spectral signature of fast shocks. I. Low-density model grid. ApJS 1996; 102: 161.
[16] Ferruit P, Binette L, Sutherland RS, Pecontal E. Modeling extragalactic bowshocks. I. The model. A&A 1997; 322: 73-85.
[17] Groves BA, Dopita MA, Sutherland RS. Dusty, radiation pressure-dominated photoionization. I. Model description, structure, and grids. ApJS 2004; 153: 9-73.
[18] Dopita MA, Binette L. Non-equilibrium modelling of the structure and spectra of shock waves. In: IAU Symposium on Supernova remnants and their x-ray emission, editors Danziger J, Gorenstein P. 1983; 101: 221.
[19] Dopita MA, Binette L, Dodorico S, Benvenuti P. Radiative shock-wave theory. I. Chemical abundance diagnostics and galactic abundance gradients. ApJ 1984; 276: 653.
[20] Dopita MA, Binette L, Tuohy IR. Radiative shock wave theory: III. The nature of the optical emission in young supernova remnants. ApJ 1984; 282: 142.
[21] Raymond JC, Hester JJ, Cox D, Blair WP, Fesen RA, Gull TR. Spatial and spectral interpretation of a bright filament in the Cygnus loop. ApJ 1988; 324: 869.
[22] Morse JA, Blair WP, Dopita MA, Hughes JP, Kirshner RP, Long KS, *et al*. Hubble Space Telescope observations of oxygen-rich supernova remnants in the Magellanic Cloud. I. Narrow-band imaging of N132D in the LMC. AJ 1996; 112: 509.
[23] Gonçalves DR, Mampaso A, Corradi RLM, Perinotto M, Riera A, Lopez-Martin L. K4-47: a planetary nebula excited by photons and shocks. MNRAS 2004; 355: 37-43.
[24] Riera A, Binette L, Raga AC. Shock excitation of the knots of Hen3-1475. A&A 2006; 455: 203-213.
[25] Shapiro PR, Kang H. Hydrogen molecules and the radiative cooling of pregalactic shocks. ApJ 1987; 318: 32.
[26] Shapiro PR, Clocchiatti A, Kang H. Magnetic fields and radiative shocks in protogalaxies and the origin of globular clusters. ApJ 1992; 389: 269.
[27] Fadeyev YA, Gillet D. The structure of radiative shock waves. II. The multilevel hydrogen atom. A&A 2000; 354: 349.
[28] Fokin AB, Massacrier G, Gillet D. Radiative cooling of shocked gas in stellar atmospheres. II. Self-consistent LTE shock wake model with Fe lines and H continua contributions. A&A 2004; 420: 1047.
[29] Chevalier RA. The evolution of supernova remnants. I. Spherically symmetric models. ApJ 1974; 188: 501.
[30] Preite-Martinez A. Late-stage evolution of a supernova remnant: the structure of the dense shell. A&A 1981; 96: 283.
[31] Lesaffre P, Chièze J-P, Cabrit S, Pineau des Forets G. Temporal evolution of magnetic molecular shocks. I. Moving grid simulations. A&A 2004; 427: 147.
[32] Lesaffre P, Chièze J-P, Cabrit S, Pineau des Forets G. Temporal evolution of magnetic molecular shocks. II. Analytics of the

steady state and semi-analytical construction of intermediate ages. A&A 2004; 427: 157.
[33] Lesaffre P. Time-dependent analytic solutions of quasi-steady shocks with cooling. 2007; Geophysical and Astrophysical Fluid Dynamics 2006 ; vol.100; Issue 3; 265-278.
[34] Walder R, Folini D. Knots, filaments, and turbulence in radiative shocks. A&A 1998; 330: L21.
[35] Folini D, Walder R. Supersonic turbulence in shock-bound interaction zones: I. Symmetric settings. A&A 2006; 459: 1.
[36] Innes DE, Giddings JR, Falle SAEG. Dynamical models of radiative shocks – II. Unsteady shocks. MNRAS 1987; 226: 67.
[37] Chevalier RA, Imamura JN. Linear analysis of an oscillatory instability of radiative shock waves. ApJ 1982; 261: 543.
[38] Smith MD, Rosen A. The instability of fast shocks in molecular clouds. MNRAS 2003; 339: 133.
[39] Innes DE. Dynamics and spectra of magnetically cushioned radiative shocks. A&A 1992; 256: 660.
[40] Landau LD, Lifshitz EM. Mécanique des fluides. MIR: Moskow 1971.
[41] Preite-Martinez A, Fusco-Femiano R. The energy-diameter relation of supernova remnants in the Large Magellanic Cloud. A&A 1986; 157: 6.
[42] Hester JJ, Raymond JC, Blair WP. The Balmer-dominated northeast limb of the Cygnus loop supernova remnant. ApJ 1994; 420: 721-745
[43] Blair WP, Sankrit R, Raymond JC. Hubble Space Telescope imaging of the primary shock front in the Cygnus loop supernova remnant. AJ 2005; 129: 2268-2280
[44] Raymond JC, Ghavamian P, Sankrit R, Blair WP, Curiel S. Far-ultraviolet spectra of a nonradiative shock wave in the Cygnus loop. ApJ 2003; 584: 770-781
[45] Binette L, Dopita MA, Tuohy IR. Radiative shock-wave theory. II. High-velocity shocks and thermal instabilities. ApJ 1985; 297: 476
[46] Cox DP. Cooling and evolution of a supernova remnant. ApJ 1972; 178: 159
[47] Tenorio-Tagle G, Rozyczka M, Franco J, Bodenheimer P. On the evolution of supernova remnants. II - Two-dimensional calculations of explosions inside pre-existing wind-driven bubbles. MNRAS 1991; 251: 318
[48] Cui W, Cox DP. Two-temperature models of old supernova remnants with ion and electron thermal conduction. ApJ 1992; 401: 206-219
[49] Blondin JM, Lundqvist P. Formation of the circumstellar shell around SN 1987A. ApJ 1993; 405: 337
[50] Chevalier RA, Dwarkadas VV. The Presupernova H II Region around SN 1987A. ApJ 1995; 452: L45
[51] Borkowski K, Szymkowiak AE, Blondin JM, Sarazin CL. A circumstellar shell model for the Cassiopea A supernova remnant. ApJ 1996; 466: 866
[52] Truelove JK, McKee CF. Evolution of nonradiative supernova remnants. ApJS 1999; 120: 299
[53] Dwarkadas VV. The evolution of supernovae in circumstellar wind bubbles. I. Introduction and one-dimensional calculations. ApJ 2005; 630: 892
[54] Dwarkadas VV. The evolution of supernovae in circumstellar wind bubbles. II. Case of a Wolf-Rayet star. ApJ 2007; 667: 226